\documentclass[twocolumn]{aastex6}

\usepackage{amsmath}
\usepackage{amssymb}

\received{October 12, 2016}
\accepted{December 29, 2016}

\shorttitle{The Unbiased Decorrelation Timescale in AGNs from SFs}
\shortauthors{Szymon~Koz{\l}owski}

\begin{document}

\title{A Method to Measure the Unbiased Decorrelation Timescale of the AGN Variable Signal from Structure Functions}

\author{Szymon~Koz{\l}owski}

\affil{Warsaw University Observatory\\Al. Ujazdowskie 4\\ 00-478 Warszawa\\Poland}
\email{simkoz@astrouw.edu.pl}


\begin{abstract}
A simple, model-independent method to quantify the stochastic variability of active galactic nuclei (AGNs) 
is the structure function (SF) analysis. If the SF for the timescales shorter
than the decorrelation timescale $\tau$ is a single power-law and for the longer ones becomes flat (i.e., the white noise), 
the auto-correlation function (ACF) of the signal can have the form of the power exponential (PE). 
We show that the signal decorrelation timescale can be measured
directly from the SF as the timescale matching the amplitude 0.795 of the flat SF part (at long timescales), 
and only then the measurement is independent of the ACF PE power.
Typically, the timescale has been measured at an arbitrarily fixed SF amplitude, but as we prove, this approach provides biased results because the AGN
SF/PSD slopes, so the ACF shape, are not constant and depend on either the AGN luminosity and/or
the black hole mass. In particular, we show that using such a method for the simulated SFs that include
a combination of empirically known dependencies between the AGN luminosity $L$ and both
the SF amplitude and the PE power, and having no intrinsic $\tau$--$L$ dependence, 
produces a fake $\tau \propto L^\kappa$ relation with $0.3\lesssim \kappa \lesssim 0.6$, that otherwise is expected from theoretical works ($\kappa \equiv 0.5$).
Our method provides an alternative means for analyzing AGN variability to the standard 
SF fitting. The caveats, for both methods, are that the light curves must be sufficiently long (several years rest-frame) and the ensemble
SF assumes AGNs to have the same underlying variability process.
\end{abstract}



\section{Introduction}

AGNs are known to be variable sources at all wavelengths 
(e.g., \citealt{1993ARA&A..31..717M, 
2004ApJ...601..692V, 
2005ApJ...618..108B, 
2005MNRAS.359.1469M, 
2010ApJ...721.1014M, 
2011ApJ...743..171A, 
2016ApJ...817..119K, 
2016A&A...593A..55V}), 
but the exact process leading to variability is still unknown, although simulations of accretion disk instabilities (\citealt{1998ApJ...504..671K}) have the closest variability 
pattern to observations in the optical bands (\citealt{1995ApJ...441..354C,2004ApJ...601..692V,2016ApJ...826..118K}). What is known, however, is that a typical AGN variability is of the stochastic 
nature (e.g., \citealt{2009ApJ...698..895K,2013ApJ...765..106Z,2013A&A...554A.137A,2016ApJ...826..118K}).
This is frequently quantified by means of the power spectral density (PSD) that
on the low frequencies shows a flat spectrum (the white noise; $\rm PSD\propto\nu^{0}$) 
and on the high frequencies appears to follow the red noise ($\rm PSD\propto\nu^{-2}$)
or even steeper dependence ($\rm PSD\propto\nu^{-3}$; e.g., \citealt{2011ApJ...743L..12M,2015MNRAS.451.4328K,2016A&A...585A.129S}). 

A similar method of quantifying the AGN variability is the structure function (SF) analysis (e.g., \citealt{1984ApJ...284..126S,1985ApJ...296...46S,1992ApJ...396..469H,1996ApJ...463..466D,2001ApJ...555..775C,2010MNRAS.404..931E,2012ApJ...753..106M,2016ApJ...826..118K}).
For a given time interval $\Delta t$ (also called the time lag), all pairs of points are identified and then the rms 
of the magnitude differences is calculated.
Typically SF, which measures the square root of the rms as a function of the time lag,  
at short time lags can be described as a single power-law (SPL) with a slope
of $\gamma\approx 0.5$ in optical-IR bands, corresponding to the PSD SPL slope of $\alpha=-2$ (e.g., \citealt{2001ApJ...555..775C,2012ApJ...753..106M,2016ApJ...817..119K,2016ApJ...826..118K}) and on the long time lags it flattens to the SPL slope of $\gamma=0$. 
The time lag at which the SF changes slope is called the decorrelation timescale $\tau$ (or the break timescale, or the decorrelation frequency for PSD), 
because for the short time lags the data points are correlated and 
for the longer ones they become uncorrelated.

It is of high interest to study the dependence of the decorrelation timescale 
on the physical parameters of AGNs, such as the black hole mass, the luminosity and/or the Eddington ratio, or its correlation with the 
dynamical, thermal, and/or viscous timescales in accretion disks (e.g., \citealt{1989MNRAS.239..289S,2001ApJ...555..775C,2006ASPC..360..265C,2008NewAR..52..253K,2009ApJ...698..895K,2014ApJ...795....2E}).
But how one does actually measure it? Typically it has been estimated as the time lag at which the SF reaches a certain, arbitrarily
selected SF amplitude. We will show that this is generally an incorrect procedure (although the only available for short light curves), because
the variability process changes with the changing black hole mass and/or the luminosity (\citealt{2016A&A...585A.129S,2016ApJ...826..118K}),
and also the SF amplitude is correlated with the luminosity. As we will show this procedure is leading to a fake relation $\tau \propto L^\kappa$ with $0.3\lesssim \kappa \lesssim 0.6$,
that is otherwise expected from the theory of accretion disks ($\tau \propto L^{0.5}$; e.g., \citealt{2002apa..book.....F,2010ApJ...721.1014M}).

In this paper, we present a method that under certain conditions (discussed in Section~\ref{sec:discussion}) produces a correct measurement of
the decorrelation timescale. If the auto-correlation function (ACF) of the stochastic process is of the power exponential (PE) form 
(which is a reasonable assumption, as explained in Section~\ref{sec:var}),
one can measure the decorrelation timescale directly from the data via the rest-frame time lag at which SF reaches the amplitude 0.795 of the 
flat SF part at long timescales. As we will show this is an unbiased measure of the decorrelation timescale, because it always returns
the the actual decorrelation timescale (and not a biased fraction of it).
One can obviously fit the SF to obtain the decorrelation timescale, however, the SF time lag bins are not independent 
producing the problems described in \cite{2010MNRAS.404..931E}.

In Section~\ref{sec:var} we describe the AGN variability, while in Section~\ref{sec:discussion} we discuss various problems related to the SF measurements and interpretations. 
The paper is summarized in Section~\ref{sec:summary}.


\section{Description of Variability}
\label{sec:var}

A light curve $y_i$ composed of $i=1, \dots, N$ points, measured at times $t_i$, can be represented as a sum of the signal $s_i$ and the noise $n_i$, i.e., $y_i=s_i+n_i$
(e.g., \citealt{1981ApJS...45....1S, 1982ApJ...263..835S, 1989ApJ...343..874S,1992ApJ...398..169R,1992ApJ...385..404P,1992ApJ...385..416P}). Empirically, from a light curve we know only $y_i$
and we do not know directly $s_i$. 
We can study the general properties of the true signal $s_i$ from the data $y_i$ using the covariance function, where we shift the copy of our light curve in time by the time difference (or the time lag)
$\Delta t=t_i-t_j$ and the $j$th index is for the copied light curve

\begin{equation}
\label{eq:sftheory}
{\rm cov}(y_i, y_j) \equiv {\rm var}(y_i) - V(y_i, y_j),
\end{equation}
\noindent
where
\begin{eqnarray}
\label{eq:covgen}
{\rm cov}(y_i, y_j) &\equiv&  \langle( y_i-\langle y \rangle)(y_j-\langle y\rangle) \rangle,\\
{\rm var}(y_i) &\equiv& \langle (y_i-\langle y \rangle)^2 \rangle,\\
V(y_i, y_j) &\equiv& \frac{1}{2}\langle(y_i-y_j)^2\rangle.
\end{eqnarray}
\noindent
The covariance of the light curve with itself is the variance var$(y_i$), $V(y_i, y_j)$ is the theoretical structure function, and $\langle \rangle$
is the summation over all $ij$ pairs in a narrow $\Delta t$ range, divided by the number of such pairs. The theoretical
SF is related to typically reported SFs via $SF=\sqrt{2V}$ (in units of magnitude, that have more natural interpretation).

From the definition and properties of the covariance, we can link the data to the signal via (from Equation~\ref{eq:sftheory})
\begin{eqnarray}
V(y_i, y_j) &=& {\rm var}(s_i) + {\rm var}(n_i) - {\rm cov}(s_i, s_j) - {\rm cov}(n_i, n_j) = \nonumber \\
            &=& \sigma_s^2 + \sigma_n^2 - {\rm cov}(s_i, s_j),
\label{eq:SFtheory}
\end{eqnarray}
\noindent
where ${\rm var}(s_i)\equiv\sigma_s^2$, ${\rm var}(n_i)\equiv\sigma_n^2$ (both the signal and noise are assumed to have the Gaussian properties),
and ${\rm cov}(s_i, n_i)={\rm cov}(n_i, n_j)\equiv 0$ because
the data are assumed here to be uncorrelated with the noise, and the noise is assumed to be uncorrelated with itself.
It is also important to note that the process leading to variability must be stationary, because only then the variances and means do not change with time.
The covariance function of the signal is related to the auto-correlation function as $ACF(\Delta t)\equiv{\rm cov}(s_i, s_j)/\sigma_s^2$.
The measured SF is then
\begin{equation}
SF_{\rm OBS}(\Delta t) = \sqrt{2\sigma_s^2(1-ACF(\Delta t))+2\sigma_n^2}.
\end{equation}

After subtracting the noise term ($2\sigma_n^2$) we have the true SF due to the variable signal only
\begin{eqnarray}
SF(\Delta t) &=& \sqrt{2\sigma_s^2(1-ACF(\Delta t))} = \nonumber \\
			&=& SF_\infty\sqrt{1-ACF(\Delta t)},
\label{eq:simone}
\end{eqnarray}
\noindent
where $SF_\infty=\sqrt{2}\sigma_s$ is the SF amplitude at timescales much longer than the decorrelation timescale 
(\citealt{2001ApJ...555..775C,2010ApJ...721.1014M,2010MNRAS.404..931E,2015MNRAS.451.4328K}). 
Throughout this manuscript we will be discussing the noise-subtracted SFs.

We are interested here in the ACF that has a form of the power exponential (PE)
\begin{equation}
ACF(\Delta t)=\exp\left[-\left(\frac{|\Delta t|}{\tau}\right)^\beta\right],
\end{equation}
\noindent
where $0<\beta\leq 2$ (e.g.; \citealt{2013ApJ...765..106Z}), because it naturally produces an SF that has one SPL slope below the decorrelation timescale and another one
(flat SF) for the longer timescales, a pattern observed in AGN SFs. This can be quantified by expanding the ACF into a Taylor series, 
where the only non-negligible terms for $|\Delta t|\ll\tau$ are $1-(|\Delta t|\tau^{-1})^{\beta}$, so the SF becomes 
an SPL of the form $SF=SF_\infty(|\Delta t|\tau^{-1})^{\frac{\beta}{2}}$. For $|\Delta t|\gg\tau$, SF becomes simply $SF\equiv SF_\infty$. 

By setting the PE power to $\beta\equiv1$, the ACF becomes the one for the damped random walk (DRW) model (\citealt{2009ApJ...698..895K,2010ApJ...708..927K,2010ApJ...721.1014M,2011ApJ...728...26M,2012ApJ...753..106M,
2011AJ....141...93B,2012ApJ...760...51R,2011ApJ...735...80Z,2013ApJ...765..106Z,2016ApJ...819..122Z}),
which is the simplest of a broader class of continuous-time autoregressive moving average (CARMA) models (\citealt{2014ApJ...788...33K}). 
DRW is nowadays frequently used to model individual AGN light curves, although the PE power seems to be $\beta>1$ for 
bright AGNs and/or massive black holes (\citealt{2016A&A...585A.129S,2016ApJ...826..118K}), causing biases in the measured DRW parameters (\citealt{2016MNRAS.459.2787K}).
Also \cite{2014MNRAS.439..703G}, by using the slepian wavelet variance method, identified a PSD break at short time scales and
concluded that DRW maybe too simplistic to describe AGN variability.

\begin{figure}
\centering
\includegraphics[width=8.3cm]{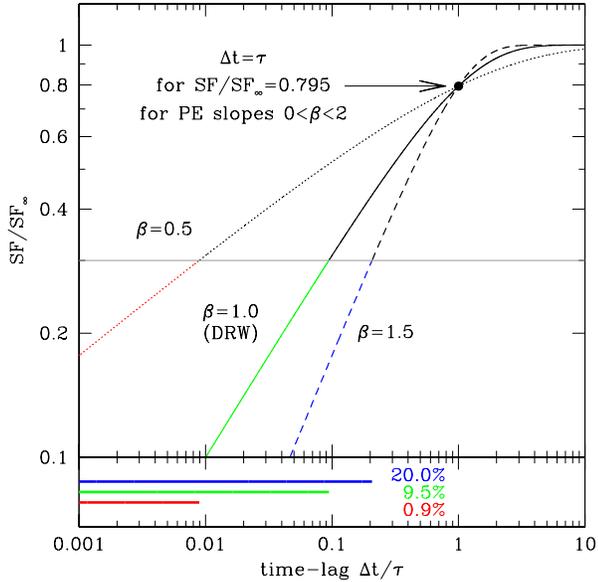}
\caption{Structure functions (Eqn.~(\ref{eq:simone})) corresponding to three ACFs with the PE power $\beta=0.5$, 1.0 (DRW), and 1.5 (from left to right). 
The only unbiased measure of the true decorrelation timescale is for the variability amplitude of $0.795~SF_\infty$.
If the timescale is measured at a different amplitude, in this example $0.3~SF_\infty$ (the gray horizontal line), 
for $\beta=0.5$ (1.0, 1.5), we in fact measure 0.9\% (9.5\%, 20\%) of the true decorrelation timescale (the bottom inset
shows projections of the three SFs below $0.3~SF_\infty$ onto the time lag axis).}
\label{fig:SF1}
\end{figure}

\subsection{The Method}

It is straightforward to prove that for $\Delta t=\tau$, SF is an unbiased measure (in terms of the underlying process) of the decorrelation timescale, because
the exponent then does not depend on $\beta$ and all $0<\beta<2$ SFs cross at the same point (Figure~\ref{fig:SF1}).
The amplitude of this point is $SF=SF_\infty\sqrt{1-\exp(-1)}=0.795~SF_\infty$. This simply means that once 
the measured SF reaches the flat part ($SF_\infty$) one can just read off the decorrelation timescale 
from the SF curve and it will be correct for the case of PE ACF regardless of the power.


\section{Discussion}
\label{sec:discussion}

Measuring either the SF amplitudes at a fixed timescale (e.g., \citealt{2004ApJ...601..692V,2010ApJ...714.1194S,2014ApJ...784...92M,2016ApJ...817..119K}) 
or the timescales at the fixed SF amplitude (e.g., \citealt{2015ApJ...798...89F,2016arXiv161103082C}) are going to provide biased results because the AGN
SF slopes at short time lags (or the PSD slopes at high frequencies) are not constant and appear to depend on either the luminosity and/or
the black hole mass (\citealt{2016A&A...585A.129S,2016ApJ...826..118K}). If the data are short and/or the break in the SF is not present, however,
this is the only justified procedure to be used.

The AGN variability amplitude is known to be anti-correlated with the luminosity (e.g., \citealt{1972IAUS...44..171A,1976AJ.....81..905U,1994MNRAS.268..305H,1994A&A...291...74P,1999MNRAS.306..637G}). 
In particular, \cite{2016ApJ...826..118K} based on the SF analysis of the 9000 SDSS AGNs
showed that the SF amplitude at long timescales (the flat part) is $SF_\infty\propto L^{-(0.35\pm0.05)}$. This means that with the increase 
of brightness by one magnitude the AGN variability amplitude decreases to about 72\%. And in fact, the amplitude of the whole SF changes by this amount.

\begin{figure}
\centering
\includegraphics[width=8.3cm]{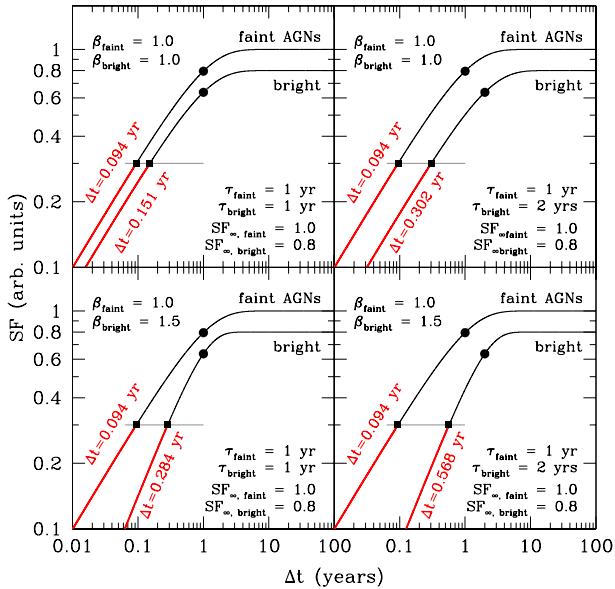}
\caption{Biases in the decorrelation timescale due to different stochastic processes and the method of measurement. The filled large dots
mark the decorrelation timescale at $0.795~SF_\infty$, while the filled squares show the timescales at the fixed SF amplitude of $0.3~SF_\infty$ (the gray line).
See Section~\ref{sec:discussion} for a detailed discussion.}
\label{fig:SF3}
\end{figure}

Measuring the decorrelation timescale at a fixed SF amplitude (below $0.795~SF_\infty$) introduces a bias, because for fainter 
AGNs with higher variability, the measured decorrelation timescale will appear shorter than the one for the brighter ones, 
even for the same intrinsic decorrelation timescale (Figure~\ref{fig:SF3}, top-left panel).
In this example, we measure the time lag at $0.3~SF_\infty$ (the gray horizontal line). For the 
faint AGNs (that have set $SF_\infty=1.0$ units) we measure 0.094 of the true decorrelation timescale, while for the brighter ones 0.151 (with set $SF_\infty=0.8$ units).
In other words, when SF is decreasing (along y-axis in Figure~\ref{fig:SF3}) because of the increasing $L$, 
this can be interpreted as a fake increase of $\tau$ (with the increasing $L$) when measuring it at a constant SF level.

While there exist an empirical evidence that the decorrelation timescale does not or weakly depend on the AGN luminosity but rather on the black hole mass, 
$\tau\propto L^{-(0.05\pm0.17)}M^{(0.38\pm0.15)}$ from \cite{2016ApJ...826..118K},
the theoretical predictions point to the form $\tau\propto L^{0.5}$ (\citealt{2002apa..book.....F}). In the top-right panel of Figure~\ref{fig:SF3}, we show what would happen if the 
decorrelation timescale had a positive correlation with the luminosity, namely the brighter the AGN the longer the timescale.

\cite{2016A&A...585A.129S} showed that the PSD slope steepens with the 
increasing black hole mass, and \cite{2016ApJ...826..118K} showed that the SF slope ($\gamma\equiv\beta/2$) steepens with the increasing luminosity as $\beta\propto L^{(0.10\pm0.03)}$.
In the bottom panels of Figure~\ref{fig:SF3}
we include this effect. This causes another bias because the measured time lag at $0.3~SF_\infty$ increases additionally
for bright AGNs. When using the empirically measured  relations $SF_\infty \propto L^{-0.35}$ and $\beta\propto L^{0.1}$, the measurement of the 
timescale at a fixed amplitude (below $0.795~SF_\infty$) produces an artificial relation $\tau \propto L^\kappa$ with $0.3\lesssim \kappa \lesssim 0.6$ 
that is otherwise expected from the theoretical standpoint, 
namely $\tau\propto L^{0.5}$, and the power of this artificial relation depends on what SF amplitude the $\tau$ measurement is made.

While it is not the goal of this paper to evaluate the biases of the SF amplitude at a fixed timescale, it is easy
to decipher from Figure~\ref{fig:SF3} what they would be. If all AGN variability was due to the same process (which is not the case)
and the timescale were independent on luminosity (which appears to be the case), the measurement of the SF amplitude would be correct
and the amplitude ratio from the bright and faint AGNs would correspond to the ratio of $SF_\infty$ for these objects (Figure~\ref{fig:SF3}, top-left panel).
If we added a theoretical positive correlation of the timescale with the luminosity, the SF amplitude at a fixed 
timescale would further decrease (Figure~\ref{fig:SF3}, top-right panel). Additional decrease will be observed for brighter 
AGNs because of the steepening of the SF slope (bottom panels of Figure~\ref{fig:SF3}).
This means that one should seek a relation of $SF_\infty$ with the physical AGN parameters and not an arbitrarily selected SF amplitude below the decorrelation timescale, that will be biased.

Obtaining a meaningful SF from a single AGN light curve that typically is short and not well sampled is problematic, if not impossible.
\cite{2010MNRAS.404..931E} have already studied and discussed various problems regarding this topic. In particular, they 
investigate the impact of data sampling and gaps, as well as data length on the SF measurements. 
The most interesting finding is that for light curves with no intrinsic decorrelation timescales (featureless PSD),
breaks will appear in the SFs of almost all light curves, and they provide a rough guide at what timescales they should appear 
as a function of the experiment length and the PSD slope (their Figure~5). While for all considered types of samplings (dense, sparse, with/without gaps)
the short time lag SF part appears to be nearly independent of the sampling, the SFs differ in shape after the spurious break.

\begin{figure*}
\centering
\includegraphics[width=5.7cm]{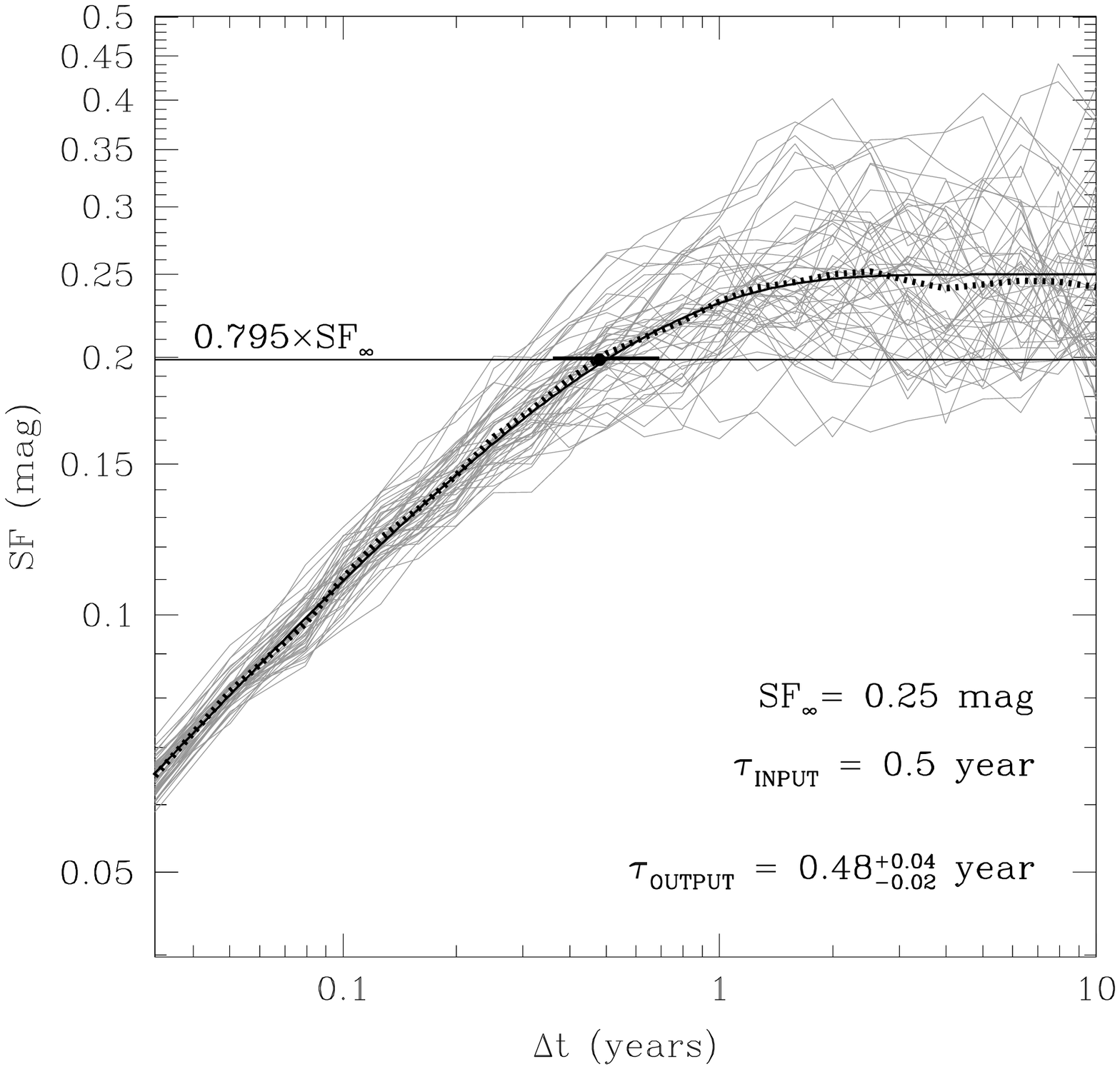}
\includegraphics[width=5.7cm]{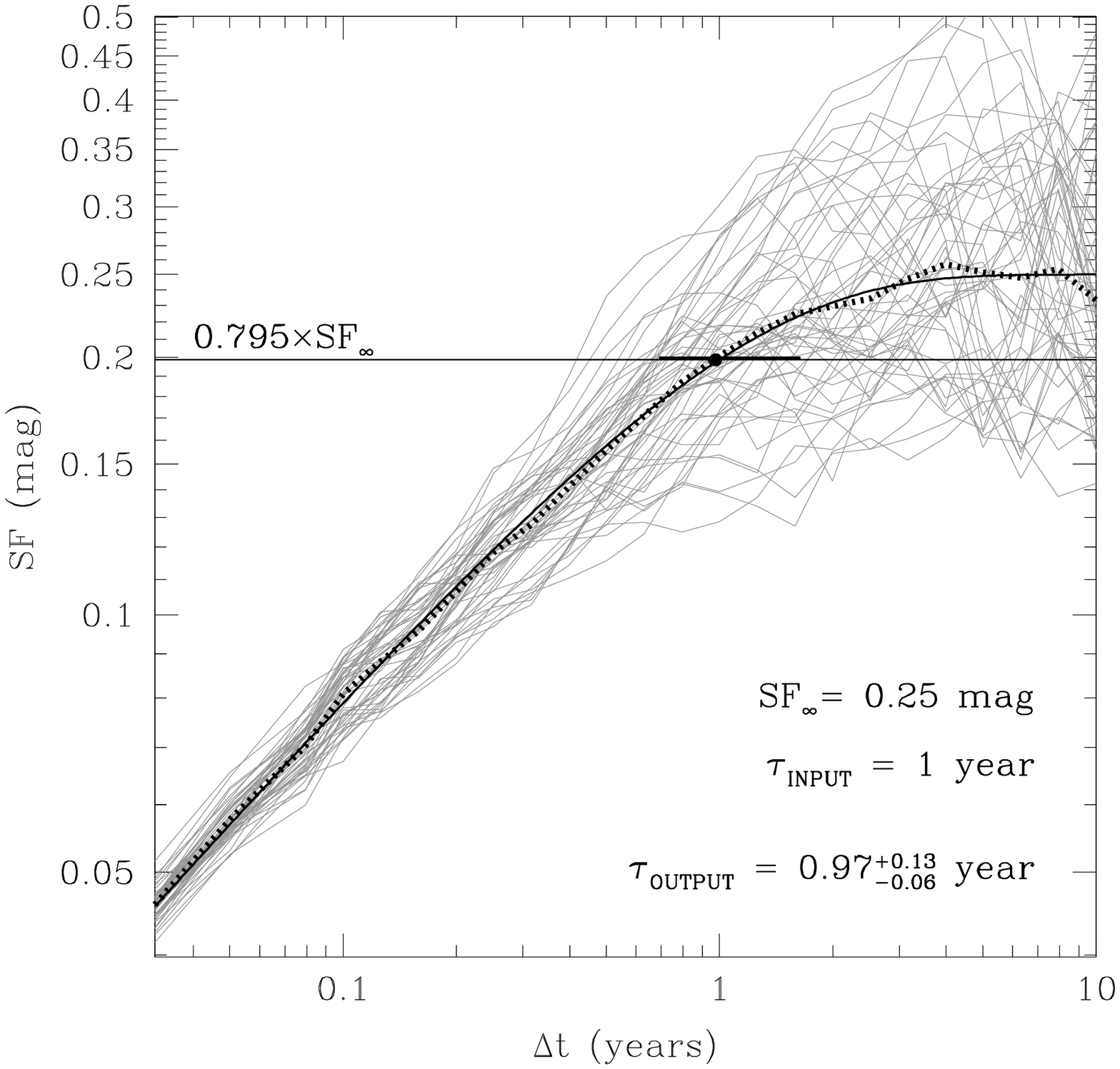}
\includegraphics[width=5.7cm]{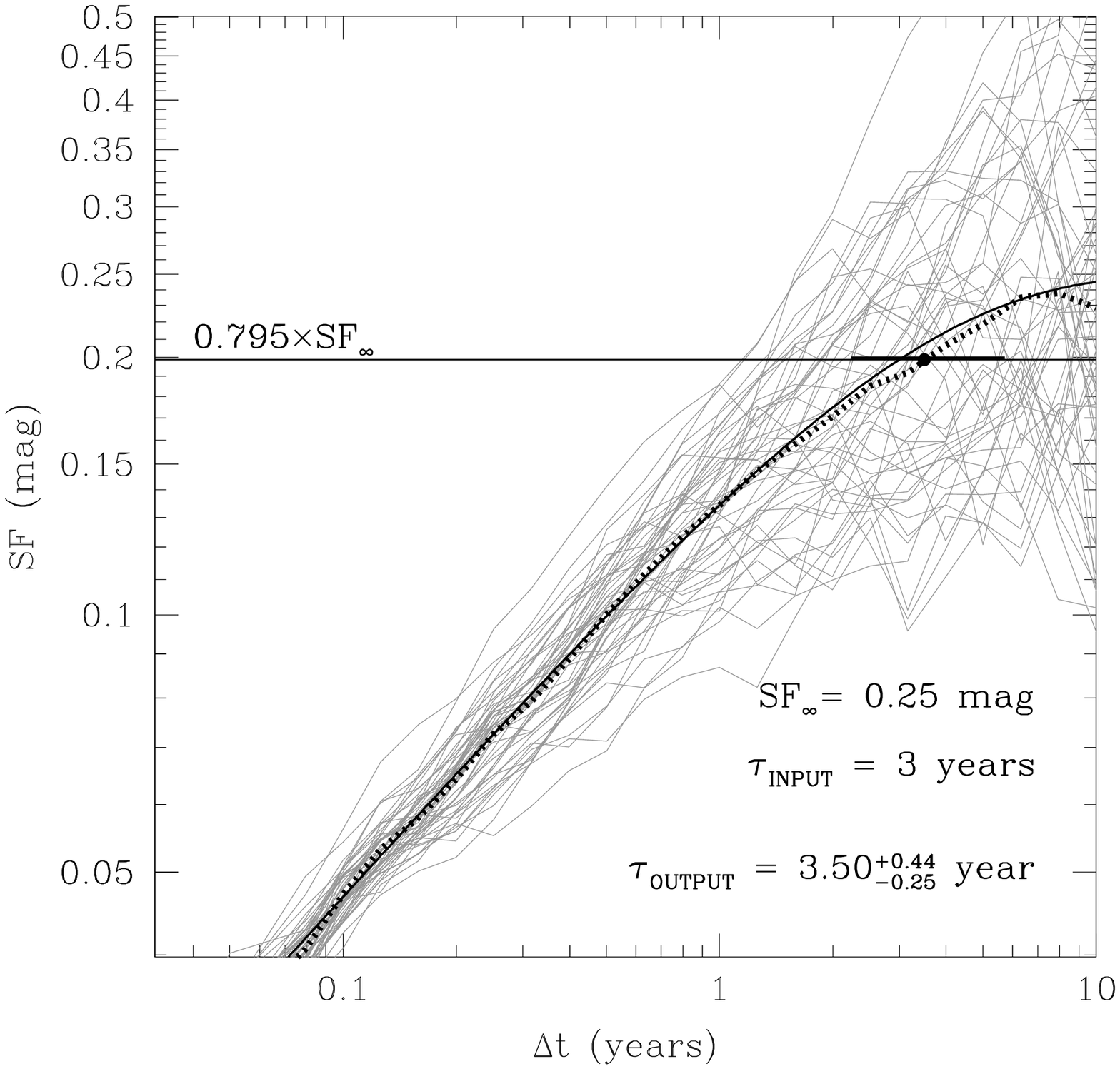}
\caption{Structure functions for 50 simulated AGN light curves (gray lines) for the same DRW process with $SF_\infty=0.25$ mag and $\tau_{\rm INPUT}=0.5$ year (left panel),
1 year (middle), and 3 years (right). The input SF for the DRW process is shown as the black solid line and the ensemble SF is shown as the black dotted line. 
\cite{2010MNRAS.404..931E} already shown that individual SFs ``suffer'' from wiggles and/or breaks that are due to the light curve length and cadence. Infinitely long and well-sampled light curves
would asymptotically produce the input SFs. A similar effect occurs when one merges a number of individual SFs (the ensemble SF),
however, it is not clear if AGNs with similar physical parameters should have the same process leading to variability (although this is commonly assumed). 
The measured decorrelation timescale $\tau_{\rm OUTPUT}$ is 
estimated at $0.795~SF_\infty$ (marked with dot) and can be well-determined from ensemble SFs, provided the data are sufficiently long to constrain $SF_\infty$.
The horizontal error bar shows the asymmetric one side dispersions, while the reported uncertainties are these dispersions divided by $\sqrt{25}$ (for each side separately).}
\label{fig:SF2}
\vspace{0.1cm}
\end{figure*}

To explore some of these problems, we simulate three sets of 50 AGN light curves spanning 5000 days (13.7 years) with the same process having $\beta=1.0$, $SF_\infty=0.25$~mag, 
and for the decorrelation timescales of $\tau=0.5$, 1, and 3 years,
sampled every 10 days, so having 500 data points (using the prescription from \citealt{2010ApJ...708..927K}). 
For every light curve, we calculate its SF (Figure~\ref{fig:SF2}, thin gray lines).
The SF for the input process is shown as the thick black line in Figure~\ref{fig:SF2}. It is obvious that each individual SF
differs from the input SF, because of the data sampling and due to different light curve realizations of the same process. We calculate the ensemble SF
from the 50 light curves that is shown as the dotted black line in Figure~\ref{fig:SF2}. It closely resembles the input SF and we show
that the measurement of $\tau$ at $0.795~SF_\infty$ is adequate (as indicated by the uncertainties). 
Note, however, we assumed here a simplification by using the exact same process for all 50 light curves (identical process parameters, but different light curve realizations).
It is not clear if this assumption holds for the variability processes for a collection of true AGNs with similar physical parameters,
although this is what is commonly assumed. 

While this question still awaits to be answered, \cite{2008AIPC.1082..282M} show that ensemble SFs from two-epoch data provide quantitatively similar results to those based on light curves with many epochs.

Another potential problem is mentioned by \cite{2010MNRAS.404..931E}, who argued that fitting a model to the SFs is an intrinsically incorrect procedure because the time lag bins are not independent,
the SF uncertainties appear too small, and the bootstrap method yields statistically meaningless SF error bars 
(these problems were also identified and discussed in \citealt{2016ApJ...826..118K}). We provide here a method of determination of the decorrelation timescale 
that is not based on SF fitting, so once the flat part of the SF can be identified, $\tau$ can be just ``read off'' from the SF at $0.795~SF_\infty$ level.
In practice, however, reaching the $SF_\infty$ level may be problematic, because one needs to collect many light curves that are several years long in rest-frame,
so for distant AGNs meaning plausibly decades.
As already mentioned, also the assumption that an ensemble of light curves for many AGNs can be treated as representative for the group,
has not been verified. It is plausible that AGNs with similar or identical physical parameters (the BH mass and luminosity) will have 
variability that is due to different processes, so ensemble variability studies may not be valid.

\cite{2011ApJ...730...52K} proposed a sophisticated method of analyzing individual AGN light curves with a mixture of DRW processes, 
and pointed out that such a mixture can result in a range of PSD slopes.
It is likely, however, that most near-future individual light curves will be either short or not well sampled to enable secure 
determination of the model parameters for large AGN samples, so ensemble SFs will be a must (although see the caveats from the previous paragraph).


\section{Summary}
\label{sec:summary}

In this paper, from basic properties of the covariance of the variable signal in the data, we derived a method of measurement the decorrelation timescale for AGN light curves
that always provides the actual and process-independent value. It is valid for SFs that at short time lags show a single power-law behavior 
and on the long ones appear to be flat, hence the ACF of the process can be of the power exponential type. The decorrelation timescale 
should be measured at $0.795$ of the SF amplitude at the long timescales (after the photometric noise is removed). We also showed that 
when using the empirically established relations $SF_\infty \propto L^{-0.35}$ and $\beta\propto L^{0.1}$, the measurement of the 
timescale at a fixed SF amplitude (below $0.795~SF_\infty$) produces an artificial non-existing relation, 
$\tau \propto L^\kappa$ with $0.3\lesssim \kappa \lesssim 0.6$ (e.g., $\kappa=0.4$ found by \citealt{2016arXiv161103082C}), 
that is otherwise expected from the theory of accretion disks (i.e., $\kappa \equiv 0.5$).

While individual SFs for typical AGN light curves, that are short and sparsely sampled, are rarely meaningful (\citealt{2010MNRAS.404..931E}),
we showed that ensemble SFs from many AGNs would yield reliable decorrelation timescales for a whole class (having assumed identical
variability parameters for individual objects). This is of particular importance
because deep, large, optical sky surveys aiming at variability (such as (in the alphabet order) Catalina/CRTS, DES, Gaia, LaSilla-Quest, LSST, OGLE, PanStarrs, and SDSS/BOSS) 
have already or will provide in the near future light curves for thousands or hundreds of thousands of AGNs. The problem that these data will face, 
however, is their length and/or cadence. AGNs are typically distant sources with
significant redshifts $z$, so the rest frame data lengths, in fact, will be shorter by a factor of $(1+z)$. Such SFs may not probe sufficiently long timescales ($SF_\infty$)
to measure the decorrelation timescale reliably. Building the ensemble SFs may remain the main tool for these data sets, because the sparseness/length of light curves 
may prevent their direct modeling (for most of the surveys; see \citealt{2016arXiv161108248K}). The caveat is that the assumption that an ensemble of light curves for many AGNs can be treated 
as representative for the group has not been verified, but is commonly assumed.

The consecutive SDSS Quasar Data Releases (e.g., \citealt{2010AJ....139.2360S,2016arXiv160806483P}) have provided increasingly rich databases of 
AGN properties that include now $\sim$280,000 black hole mass estimates, 
the luminosities, and the Eddington ratios (e.g., \citealt{2011ApJS..194...45S,2016arXiv160909489K}) distributed over a quarter of the sky, 
enabling unprecedented studies of the connection between the AGN variability and the underlying AGN physics. The forthcoming decades 
are guaranteed to bring many new and exciting developments in this field of research.


\acknowledgments

We are grateful to the anonymous referee for reading the manuscript and providing us with comments that 
improved the flow and clarity of the presented arguments.
S.K. acknowledges the financial support of the Polish National Science Center through the
OPUS grant number 2014/15/B/ST9/00093 and MAESTRO grant number 2014/14/A/ST9/00121.


\end{document}